# Electrical spin injection and accumulation in CoFe/MgO/Ge contacts at room temperature

Kun-Rok Jeon[1], Byoung-Chul Min[2], Young-Hun Jo[3], Hun-Sung Lee[1], Il-Jae Shin[2], Chang-Yup Park[1], Seung-Young Park[3] and Sung-Chul Shin[1]

[1]*Department of Physics and Center for Nanospinics of Spintronic Materials, Korea Advanced Institute of Science and Technology (KAIST), Daejeon 305-701, Korea*

[2]*Center for Spintronics Research, Korea Institute of Science and Technology (KIST), Seoul 136-791, Korea*

[3]*Nano Materials Research Team, Korea Basic Science Institute (KBSI), Daejeon 305-764, Korea*

**We first report the all-electrical spin injection and detection in CoFe/MgO/ moderately doped *n*-Ge contact at room temperature (RT), employing three-terminal Hanle measurements. A sizable spin signal of ~170 k$\Omega$ $\mu m^2$ has been observed at RT, and the analysis using a single-step tunneling model gives a spin lifetime of ~120 ps and a spin diffusion length of ~683 nm in Ge. The observed spin signal shows asymmetric bias and temperature dependences which are strongly related to the asymmetry of the tunneling process.**

# I. INTRODUCTION

The rapid evolution of electronics requires alternative technologies more than scaling down the device size, and spintronics based on the electron spins in semiconductor raises prospects for future electronics.[1-6] The electrical injection of spin-polarized electrons from ferromagnet (FM) into semiconductor (SC) and subsequent detection of the resultant spin accumulation provide a viable route for the realization of semiconductor-based spintronics.[1-6] The electrical spin injection into GaAs, InAs, or Si from FM through a spin-dependent tunnel barrier has been demonstrated using optical detection in spin lighting emitting diodes[7-10] or electrical detection in vertical/lateral (spin valve) structures.[11-18] With engineering of magnetic tunnel contacts, significant spin signals have been observed in Si using Co/NiFe/$Al_2O_3$ and Fe/$SiO_2$ tunnel contacts up to room temperature (RT).[6,19]

Recently, the *n-type* Ge in conjunction with a crystalline bcc FM/MgO(001)[20-23] has attracted much attention as a promising candidate for the efficient spin injection in terms of a high tunnel spin polarization (TSP), a small conductivity mismatch, and a negligible interdiffusion/intermixing in FM/Oxide/SC contacts. Moreover, considering a high electron mobility in Ge (at least twice higher than Si) and its weak dependence on doping concentration, Ge prospectively represents a SC channel with a long spin diffusion length.[2,24] Several important achievements[25,26] have been recently reported in the field of spin transport in Ge at low temperature, but the spin injection and detection in Ge at RT is yet to be investigated.

Here we first demonstrate the electrical spin injection in spin tunnel contacts consisting of crystalline bcc CoFe/MgO (001)/moderately doped *n*-Ge and the electrical detection of the induced spin accumulation at RT. We have analyzed the spin accumulation, spin life time, spin diffusion length in Ge from the measured spin signal, and studied their bias and temperature dependences.

## II. EXPERIMENTAL DETAILS

### A. Principle of the approach

Figure 1(a) illustrates the device geometry and measurement scheme used in the present study. We have fabricated a symmetric device consisting of five single crystalline CoFe/MgO/$n$-Ge tunnel contacts (*a-e*) spaced as shown in the inset of Fig. 1(a). The contacts *a/b/c* (30×100/20×100/20×100 μm$^2$) are used as spin injectors/extractors and also spin detectors, while the contacts *d/e* (150×100/150×100 μm$^2$) are used as references. The contacts are separated from each other more than 100 μm, which is much longer than the spin diffusion length. The magnetic easy axis of the CoFe contacts are along the [110] direction of Ge in parallel to the long axes of the contacts. The measurement scheme[6,18,19] (Fig. 1(a)) using a single contact in the three-terminal geometry provides a simple way to measure the induced spin accumulation in SC by spin injection or extraction.

When the spin-polarized electrons are injected from $FM_1$ to SC, majority spins accumulate in SC (at $x_1$, $\Delta\mu_+ = \mu_+^\uparrow - \mu_+^\downarrow > 0$); when the electrons (mostly majority-spin electrons) are extracted from SC to $FM_1$, minority spins accumulate in SC (at $x_1$, $\Delta\mu_- = \mu_-^\uparrow - \mu_-^\downarrow < 0$) as shown in Figs. 1(b). This spin accumulation induced by spin injection or extraction can be detected electrically using the same contact by means of the Hanle effect.[6,16,18,19] A transverse magnetic field ($B_\perp$) suppresses the spin accumulation in the SC (at $x_1$) via spin precession, and results in a voltage drop between $FM_1$ and $FM_2$ as a function of the applied field ($B_\perp$) (i.e. negative magnetoresistance (MR)) as depicted in Fig. 1(b). Ignoring recombination effects, the voltage drop ($\Delta V$) can be described approximately by a Lorentzian function, $\Delta V_\mp(B_\perp) = \Delta V_\mp(0)/(1+(\Omega \tau_{sf})^2)$,[27] with $\Delta V_\mp(0) = \gamma \Delta \mu_\pm(0)/(-2e)$, $\Omega = g\mu_B B_\perp/\hbar$. Here $\gamma$ is the TSP of the tunnel contact, $g$ is the Lande $g$-factor, $\mu_B$ is the Bohr magneton, and $\tau_{sf}$ is the spin lifetime. From the above relation, one can extract the spin lifetime of carriers ($\tau_{sf}$) and spin accumulation ($\Delta \mu$) in SC.

Three-terminal Hanle measurement cannot fully uncover whether the measured spin accumulation comes from the bulk SC channel[28] or the localized states (LSs) at the interface.[18] It has been argued that the observed Hanle spin signal comes from the LSs in Co/Al$_2$O$_3$/GaAs contact which have a wide depletion region and large contact resistance.[18] In contrast, the recent report[28] studying the NiFe/Al$_2$O$_3$/Cs/$n$-Si contact, which have a narrow depletion region and small contact resistance, demonstrates that the spin polarization exists in the bulk bands of the SC rather than in LSs. These studies showed that the measured spin signals are closely associated with the contact characteristics such as the width of the depletion region ($W_d$) and the resistance area (RA) product.

**B. Structural and electrical characterization**

Figure 1(c) shows *in-situ* reflective high-energy electron diffraction patterns of the MgO(2 nm) layer and CoFe(5 nm) layer after annealing at 300 °C, low-magnification and high-resolution transmission electron microscope images, and selected area electron diffraction covering the whole region of the CoFe(5 nm)/MgO(2

nm)/*n*-Ge tunnel structure. These *in-situ* and *ex-situ* structural characterizations confirm the single-crystalline nature of the tunnel structure and the in-plane crystallographic relationship of CoFe(001)[100]∥MgO(001)[110]∥Ge(001)[100], exhibiting sharp interfaces in the (001) matching planes. This crystalline tunnel structure with a 4-fold in-plane crystalline symmetry is desirable for efficient spin injection with a high TSP via the symmetry-dependent spin filtering effect of MgO(001) barrier in conjunction with bcc FM.[7,29]

Figure 1(d) shows the typical *J-V* characteristics of the CoFe(5.0 nm)/MgO($t_{MgO}$=1.5, 2.0, 2.5 nm)/*n*-Ge tunnel contacts with the electric resistivity ($\rho$) of 7.5-9.5 mΩ cm and a moderate doping concentration ($n_d$) of $2.5 \times 10^{18}$ cm$^{-3}$, well below the metal-insulator transition ($1.04 \times 10^{19}$ cm$^{-3}$), at 300 K. As shown in *J-V* curves, a rectifying behavior is gradually reduced with increasing the MgO thickness, indicating that the Schottky characteristics have been considerably suppressed. For a quantitative analysis, we have estimated the RA product (*V/J*), the Schottky barrier height (SBH, $\Phi_B$) and the depletion width ($W_d$) using the conventional *I-V-T* method. The estimated values are shown in Fig. 1(e). In this figure, we see that a thicker MgO layer effectively reduces the SBH with the cost of increase of tunnel resistance. This result is fairly consistent with the Fermi-level depinning (FLD) mechanism[21-23] in metal/insulator/Ge contacts. As a consequence, we have effectively tuned the energy-band profile of the CoFe/MgO/*n*-Ge contact by adjusting the MgO thickness (i.e. 2-nm MgO in our system) for the spin injection and detection approach in moderate doped *n*-Ge at RT.

## III. RESULTS & DISCUSSION

## A. Electrical injection and detection of spin accumulation in Ge at 300 K

The spin accumulation in the CoFe/MgO/*n*-Ge contact is measured by the voltage changes ($\Delta V$) as a function of a transverse magnetic field ($B_\perp$) at the bias voltages of $\mp$ 0.15 V in the temperature range 200-300 K. As shown in the $\Delta V$-$B_\perp$ plots (Fig. 2(a)), the tunnel contact clearly exhibits the negative MR with a Lorentzian line shape, indicating that the induced spin accumulation in Ge by spin injection or extraction is effectively detected. It is noteworthy to mention that the spin tunnel contact with a small $\Phi_B$ of 0.25 eV and a narrow $W_d$ of 12 nm enables us to observe the spin signals with both forward and reverse bias polarities in the temperature range 200-300 K.[16, 30] Albeit the significant suppression of the SBH, the still remaining Schottky barrier results in a resistive contact at low temperature and make it difficult to obtain enough $\Delta V$ signals below 200 K.

## B. Control experiment

The anisotropic MR (AMR) of the FM is negligible in our experiment, since the resistance of the FM contact is at least two orders of magnitude smaller than the tunnel resistance. The Lorentz MR (LMR) of the Ge channel cannot explain this voltage change, since the resistance of the SC increases with the applied magnetic field in the LMR effect. In order to exclude any artifacts caused by the stray field near the edges of the FM, we have conducted the control experiments using the CoFe(5 nm)/Cr($t_{Cr}$=1.5, 3.0 nm)/MgO(2 nm)/Ge tunnel contacts by inserting the non-magnetic Cr between CoFe and MgO,[6] which is effective to reduce the tunnel spin polarization without significantly changing the stray field (note that no significant changes of the structural and electrical properties were observed in the Cr-inserted tunnel contacts compared to the tunnel contact without the Cr layer; see the Appendix B). As shown in Fig. 2(b), a strong suppression of the MR signal is observed with increasing the Cr thickness ($t_{Cr}$), verifying that the observed MR signals in the CoFe/MgO/Ge contact is purely originated from the spin accumulation.

## C. Estimation of spin accumulation, spin life time, spin diffusion length, and spin polarization in Ge

Figure 2(c) shows the electrical Hanle signals ($\Delta V$) as a function of a transverse magnetic field at RT with the applied currents of $-14/+179$ μA, corresponding to $V_{\mp} = \mp 0.15$ V at $B_{\perp}=0$. The most salient feature of Fig 2(c) is clear and significant Hanle signals obtained at RT for the both conditions of spin injection/extraction ($V_{\mp}$). A remarkable spin RA product (or spin signal, $\Delta V / J$) as large as 170 $k\Omega\ \mu m^2$ is obtained across the CoFe/MgO/Ge tunnel contact for the low bias voltage ($V_{-} = -0.15$ V), which is an order of magnitude greater than that of Co/NiFe/AlO/$n$-Si contact.[6]

The estimation of the spin accumulation, spin life time, spin diffusion length, and spin polarization in Ge from the measured spin signal strongly depends on a model describing the tunneling process in the spin tunnel contacts. Taking into account the narrow $W_d$ (~12 nm) and the relatively small $RA$ of the our contact (~$3\times10^{-5}\ \Omega m^2$ at -0.15 V), two orders of magnitude smaller than that in Ref. 18, we have analyzed the measured results based on a single-step tunneling process instead of the two-step tunneling process.[18] The two-step tunneling could be possible as long as the interface and the SC bulk channel are sufficiently decoupled by a wide Schottky barrier (see Eq.

(C5) in the Appendix C). A narrow depletion region might facilitate a single-step tunneling from a Ge/CoFe to CoFe/Ge across the depletion region without loss of spin polarization. Hence, the interface and the Ge bulk channel are directly coupled, which equalizes their spin accumulation (see Eq. (C4) in the Appendix C, Fig. 5).

We have calculated the spin accumulation $\Delta\mu_+ \approx (+)2.23$ mV at the Ge interface from $\Delta\mu_+ = (-2e)\Delta V_- / \gamma_-$, using the measured Hanle signal of $\Delta V_- \approx (-)0.78$ mV. In this calculation, the TSP ($\gamma_-$) value of crystalline CoFe/MgO tunnel contact was assumed to be 0.7,[31] because the experimental data for the TSP of the CoFe/MgO/Ge contact is not available; this TSP value is likely to be a higher bound. Assuming a parabolic conduction band and a Fermi-Dirac distribution for each spin and using the calculated spin accumulation, $\Delta\mu_+ \approx (+)2.23$ meV, we have determined the associated spin polarization in the Ge, $n_\uparrow - n_\downarrow / n_\uparrow + n_\downarrow \approx (+)4.4\%$, where $n_\uparrow / n_\downarrow \approx 1.31 \times 10^{18}$ cm$^{-3}$ / $1.20 \times 10^{18}$ cm$^{-3}$ are the density of spin up/down electrons.[24] We believe that spin polarization might be larger than (+)4.4 %, since we have used a highest value of $\gamma = 0.7$.

Using a Lorentzian fit and taking an electron $g$-factor of -1.6 for the Ge, we have obtained the spin lifetime of $\tau_{sf,-} \approx 120$ ps ($V_- = -0.15$ V) in moderately doped $n$-Ge at RT. Such timescale is much smaller than the expected spin lifetime (order of a ns) of conduction electrons in moderately doped $n$-Ge from the Elliott-Yafet spin relaxation rate.[2,32,33] However, we believe that the true spin lifetime may be longer than $\tau_{sf,-} \approx 120$ ps. According to the recent report,[34] the local magnetostatic field due to the finite roughness of the FM/Oxide interface strongly reduces spin accumulation at the SC interface and artificially broadens the Hanle curve. As proven by the in-plane measurement (*M//B*), showing the inverted Hanle effect (Fig. 2(d), blue), the interfacial depolarization effect is considered as a main origin of the unexpectedly broadened Hanle curve in this system. Hence, the true spin lifetime is expected to be longer, and its temperature dependence is masked by the effect of the local magnetic field (see Fig. 2(a)).

It should be noticed that the Hanle curve has a slightly broader width for the reverse bias ($V_- = -0.15$ V, spin injection) than the forward bias ($V_+ = +0.15$ V, spin extraction). The broadening effect of Hanle curves due to the local magnetic field can be quantified using a parameter $\Delta V_{inverted}/\Delta V_{normal}$. As shown Fig. 2(d), the $\Delta V_{inverted}/\Delta V_{normal}$ is more or less the same for both reverse and forward bias. This implies that the bias dependence of the spin lifetime could be caused by other mechanisms, for example, unequal momentum scattering rates[32,33] for the injected and extracted electrons or differences in the tunneling process (see section D)

In addition, we have calculated the spin diffusion length $l_{sf} = \sqrt{D\tau_{sf}}$ in the Ge, where $D$ is the diffusion coefficient ($D \approx 38.9$ cm$^2$ s$^{-1}$ at RT estimated from the Einstein relation using the mobility ($\mu$) versus doping concentration ($n_d$) relation).[24] With $\tau_{sf,-} \approx 120$ ps, we have obtained the corresponding spin diffusion length $l_{sf,-} \approx 683$ nm at 300 K. This value is about three times larger than that of the electron spin diffusion length (230 nm) of the degenerate *n*-Si (As-doped, $\rho = 3$ m$\Omega$ cm).[6]

## D. Bias voltage dependence of spin signal

The electrical Hanle signal ($\Delta V$) and the spin RA product ($\Delta V/J$) of the CoFe/MgO/*n*-Ge contact show a strong bias dependence (Figs. 3(a),(b)): those data are significantly asymmetric with respect to the voltage polarity. The Hanle signal increases gradually with the reverse bias ($V_-<0$, spin injection), but varies slightly with the forward bias ($V_+>0$, spin extraction). The spin RA product shows a similar bias dependence as reported in the Co/NiFe/Al$_2$O$_3$/$n$-Si contact.[6]

In order to understand the asymmetric bias dependence of the spin signal (or spin RA product), we utilize the equation describing the spin signal at the Ge interface[3,18]: $\Delta V/J = \gamma_d \gamma_{i/e} r_{ch} = \gamma_d \gamma_{i/e} \rho \sqrt{D\tau_{sf}}$. Here, $\gamma_d$ is the TSP corresponding to the detection of induced spin accumulation at the Ge interface, $\gamma_{i/e}$ is the other TSP of the injected/extracted electrons, and $r_{ch}$ is the spin-flip resistance associated with the Ge bulk channel.

According to the above equation, the $\Delta V/J$ is proportional to $\gamma_d \gamma_{i/e} \sqrt{\tau_{sf}}$ at a given temperature ($T$), which depends on $V$. Using the $\Delta V/J$ values (Fig. 3(b)) and $\tau_{sf}$ values (not shown) extracted from the Lorentzian fit, we have plotted the TSP$^2$ ($\gamma_d \gamma_{i/e}$) vs. the $V$ at different temperatures to extract the bias dependence of TSP in Fig. 3(c), where the TSP$^2$ data is normalized by the maximum value at each temperature. Interestingly, TSP$^2$ becomes independent of bias voltage for $V_-<0$ (gray line in Fig. 3(c)), but decays exponentially for $V_+>0$ (black line in Fig. 3(c)). With the assumption of $\gamma_d = \gamma_{i/e}$, the variation of TSP with $V$ is then obtained as $\gamma_- \propto \gamma_o$ and $\gamma_+ \propto \gamma_o \exp(-eV_+/0.06)$. This is qualitatively similar to that of FM/I/NM (nonmagnet) tunnel contacts.[35,36] The asymmetry of TSP observed in FM/I/NM contacts is mainly due to the intrinsic asymmetry of tunneling process with respect to bias polarity[35]: the electron tunneling out of the FM originates near the Fermi-level with relatively large polarization ($V_-<0$, Fig. 5(b)), whereas the electron tunneling into the FM faces hot electron states well above the Fermi-level with significantly reduced polarization ($V_+>0$, Fig. 5(a)). Therefore, the asymmetric bias dependence of the spin signal in our system is understood in terms of the asymmetry of TSP caused by the intrinsic asymmetry in these tunneling processes.[35]

**E. Comparison of obtained spin signal with existing drift-diffusion model**

It should be noticed here that the obtained spin signal ($\Delta V / J$, red rectangle in Fig. 3(d)) for the reverse bias ($V_-<0$) is more than two orders of magnitude larger than the expected value from the single-step tunneling ($r_{s\_ss} = \gamma_d \gamma_{i/e} \rho \sqrt{D\tau_{sf}}$, red circle in Fig. 3(d)). It is tempting to explain this discrepancy using a different tunneling model. For example, the unexpected large spin signal was also found in Co/AlO/$n$-GaAs tunnel contact[18] at low temperature which was explained by the contribution of the two-step tunneling process through the LSs nearby the SC interface (e. g., interface states at the Oxide/SC, ionized impurities in the depletion region), where the LSs act as an intermediate stage for the spin injection ($V_-<0$) and absorb most of the spin polarization before it reach the SC. However, the measured spin signal also shows a large discrepancy with the spin signal estimated from the two-step tunneling ($r_{s\_ts} = \gamma_d \gamma_{i/e} r_{LS} = \gamma_d \gamma_{i/e} \tau_{sf} / e^2 N^{LS}$, with $N^{LS} \sim 5\times10^{13}$ eV$^{-1}$ cm$^{-2}$,[22] red triangle in Fig. 3(d)) The calculated spin signal from the two-step tunneling even with an optimistic spin lifetime (~1 ns) is still about one order of magnitude smaller than that of obtained spin signal (see open triangle in Fig. 3(d)). Moreover, the two-step tunneling process cannot explain the exponential increase of our spin signal (Fig. 3(e)) with the temperature decrease, as the two-step tunneling predicts only a modest increase of the spin signal with decreasing the temperature from 300 K to 200 K.

Because of the limitation of the three terminal Hanle measurements, the optical or non-local measurement of spin signals is required to unambiguously determine whether the observed spin signal in this system originates from the spin accumulation in the Ge bulk channel or LSs.

**F. Underestimation of real/local current density**

A large deviation of the obtained spin signal ($\Delta V/J$) from those estimated from a single-step tunneling model has been also reported in the tunnel contacts on moderately doped Si (magenta[28] and cyan[19] symbols in Fig. 3(d)).[19,28] It has been argued that the unexpected large spin signal ($\Delta V/J_{av}$) is mainly associated with the underestimation of real/local current density ($J_{local}$),[6] not the LSs effect. The lateral distribution of tunneling current across the tunnel contact is inhomogeneous with the variation of thickness and the composition of the tunnel barrier[6] (note that the contact resistance of CoFe/MgO/Ge is very sensitive to the MgO thickness, see Figs. 1(e),(f)). Hence, the local current density ($J_{local}$, $I/A_{local}$) which induces the spin accumulation at the contact is expected much larger than the average current density ($J_{av}$, $I/A_{geo}$) estimated from the geometrical contact area ($A_{geo}$) (see Fig. 3(f)).[6]

Using this picture, we can also explain the exponential dependence of $\Delta V/J_{av}$ on $T$ in a consistent way. The electron transport in our contacts basically consists of the tunneling (or field emission, FE) and thermionic field emission (TFE) with a SBH of 0.25 eV and a $W_d$ of 12 nm. As $T$ decreases, the TFE process is strongly suppressed (see $I$-$T$ plot in Fig. 3(e)). Hence, the electron tunneling is confined within narrow paths with a relatively thinner tunnel barrier (Fig. 3(f)), since the tunnel transmission is exponentially dependent on the thickness of barrier. This confinement results in the significant increase of the $J_{local}$ ( $J_{local} >>> J_{av}$ ) by several orders of magnitude.

## IV. SUMMARY

In conclusion, we have experimentally demonstrated the electrical spin accumulation in tunnel contacts consisting of crystalline bcc CoFe/ MgO (001)/ moderately doped *n*-Ge at RT, employing three-terminal Hanle measurements. A sizable spin signal of ~170 kΩ μm$^2$, spin polarization of ~(+)4.4 %, spin lifetime of ~120 ps, and spin diffusion length of ~683 nm are obtained at RT. We find that the asymmetric bias dependence of spin signal is strongly related to the asymmetry of tunnel spin polarization. We expect that our experimental findings will lead towards the interface engineering of FM/MgO/*n*-Ge systems for efficient spin injection and detection, and, eventually, pave a way to realize Ge-based spintronics at RT.

## ACKNOWLEDGMENTS

This work was supported by the National Research Laboratory Program Contract No. R0A-2007-000-20026-0 through the National Research Foundation of Korea funded by the Ministry of Education, Science and Technology, the KIST institutional program, and the KBSI Grant T31405 for Young-Hun Jo.

## APPENDIX A: Sample preparation

The single crystalline CoFe(5 nm)/MgO($t_{MgO}$ nm)/*n*-Ge (Sb-doped, $\rho \approx$ 7.5-9.5 mΩ cm) tunnel structures were prepared by molecular beam epitaxy (MBE) system with a base pressure better than $2\times10^{-10}$ torr. To obtain a clean and flat surface, we have conducted the cleaning procedure combined *ex-situ* chemical cleaning and *in-situ* ion bombardment and annealing process.[20] All layers were deposited by e-beam evaporation with a working pressure better than $2\times10^{-9}$ torr. We used a single crystal MgO source and rod-type CoFe with a composition of $Co_{70}Fe_{30}$. The $t_{MgO}$ -nm MgO and 5-nm thick CoFe layers were grown at 125 °C and RT, respectively, and then the samples were subsequently annealed *in-situ* for 30 min at 300 °C below $2\times10^{-9}$ torr to improve the surface morphology and crystallinity. Finally, the samples were capped by a 2-nm thick Cr layer at RT to prevent oxidation of the sample. The final sample structure was an Cr(2 nm)/CoFe(5 nm)/MgO($t_{MgO}$ nm)/*n*-Ge(001). The symmetric device consisting of five tunnel contacts with lateral sizes of $30\times100/20\times100/20\times100/150\times100/150\times100$ $\mu m^2$ was prepared by using micro-fabrication techniques (e.g., photo-lithography and Ar-ion beam etching)[22] for the electrical Hanle measurement.

## APPENDIX B: Structural and electrical characterization of Cr-inserted tunnel contacts

The control experiment to exclude the artifacts caused by the stray field should be based on a structurally and electrically identical sample except the Cr insertion layer. In order to confirm this, we have analyzed CoFe(5 nm)/Cr($t_{Cr}$=0, 1.5, 3.0 nm)/MgO(2 nm)/*n*-Ge samples by using *in-situ* reflective high energy electron diffraction (RHEED) and conventional *I-V-T* measurements for the structural and electrical characterizations, respectively.

The Cr layers of CoFe/Cr/MgO/*n*-Ge samples were grown by e-beam evaporation at RT with a working pressure better than $2\times10^{-9}$ torr. Except the insertion of Cr layer, all layers were prepared under the same growth condition described in the Appendix A. It should be noticed that the Cr layer on MgO/Ge surface was not grown layer-by-layer, because the Cr does not well wet on the MgO(001) surface due to the substantially large surface energy of Cr(001) (3.98 J/m$^2$) compared with that to the MgO(001) surface (1.16 J/m$^2$).[38,39] Thus, RHEED patterns (Fig. 4(a)) of the CoFe(001) layers (with the surface energy of 2.55 J/m$^2$)[39] grown on three dimensional Cr/MgO/Ge surface show more distinct spot patterns than the CoFe layer grown on MgO/Ge surface. However, after *in-situ* annealing at 300 °C, the surface morphology and crystallinity of the CoFe layers become comparable to each other, as exhibited by the streaky patterns in Fig. 4(a). Although chemically inhomogeneous interface might be formed at the CoFe/Cr interface during the post annealing process, it is known that the Fe grown on Cr system does not show a significant interface alloying because the binding energy of Cr layer is larger than that of Fe adatoms.[40] It is believed that interdiffusion/intermixing is not significant in this system.

The *J-V* characteristics of CoFe(5 nm)/Cr($t_{Cr}$=0, 1.5, 3.0 nm)/MgO(2 nm)/*n*-Ge tunnel contacts (Fig. 4(b)) show quasi-Ohmic behaviors for the entire contacts at RT, except for much symmetric features in the Cr-inserted tunnel contacts that might be expected due to the lower work function of Cr (4.5 eV) than CoFe (4.75 eV). Moreover, using the conventional *I-V-T* method, we have deduced Schottky barrier height (SBH) of each contact. The SBHs estimated from the slope of the Arrhenius plots $(\ln(I_R/T^2)-1/T)$ by the linear fit at reverse bias of -0.15 V (Fig. 4(c)) are 0.25, 0.23, and 0.24 eV for the Cr thickness ($t_{Cr}$) of 0, 1.5, and 3.0 nm, respectively. It indicates that the insertion of Cr layers does not affect major electrical features of the CoFe/MgO/*n*-Ge contact.

As a result, we can rule out another possible origin for the strong suppression of MR signal due to significant changes of the structural and electrical properties of the tunnel contacts by the insertion of Cr layer.

**APPENDIX C: Existing drift-diffusion model**

To examine the possibility of two-step tunneling process (or LSs effect) in our system, here we adopt a model,[18] taking into account the two-step tunneling process through LSs (e. g. interface states at the Oxide/SC, ionized impurities in the depletion region).

According to the model,[18] the spin accumulations in the Ge (LSs ($\Delta\mu_{LS}$), $n$-Ge channel ($\Delta\mu_{ch}$)) and the spin signal ($\Delta V/V$) are expressed as:

$$\Delta\mu_{LS} \approx 2e\gamma J \frac{r_{LS}(r_b + r_{ch})}{r_b + r_{LS} + r_{ch}}, \quad \Delta\mu_{ch} \approx 2e\gamma J \frac{r_{LS} r_{ch}}{r_b + r_{LS} + r_{ch}}, \tag{C1}$$

$$\frac{\Delta V}{V} \approx \frac{\gamma^2}{1-\gamma^2}\left(\frac{r_{LS}}{R_b^* + r_b}\right)\frac{r_b + r_{ch}}{r_b + r_{LS} + r_{ch}} = \frac{\gamma^2}{1-\gamma^2}\left(\frac{\tau_{sf}}{\tau_n}\right) \tag{C2}$$

with

$$\tau_{sf} \approx \tau_{sf}^{LS} \frac{N^{ch}\tau_{\rightarrow}^{LS} + (N^{ch} + N^{LS})\tau_{sf}^{ch}}{N^{ch}(\tau_{\rightarrow}^{LS} + \tau_{sf}^{LS}) + N^{LS}\tau_{sf}^{ch}}, \quad \tau_n \approx \left(1 + \frac{N^{ch}\tau_{sf}^{ch}}{N^{LS}\tau_{sf}^{ch} + N^{ch}\tau_{\rightarrow}^{LS}}\right)(\tau_{\leftarrow}^{LS} + \tau_{\rightarrow}^{LS}), \tag{C3}$$

where $R_b^* = \tau_{\leftarrow}^{LS}/(e^2 N_{3D}^{LS} d_{LS})$ is the spin-dependent tunnel resistance of the MgO layer, $r_b = \tau_{\rightarrow}^{LS}/(e^2 N_{3D}^{LS} d_{LS})$ is the bias-dependent leakage resistance at the LSs and the $n$-Ge bulk channel, and $r_{LS/ch} = \tau_{sf}^{LS/ch}/(e^2 N_{3D}^{LS/ch} d_{LS/ch})$ are the spin-flip resistances associated with these LSs and $n$-Ge bulk channel. $\tau_{sf}^{LS/ch}$, $N_{3D}^{LS/ch}$, and $d_{LS/ch}$ are the spin lifetime, density of states per unit volume and thickness of each layer. The $\tau_{\leftarrow/\rightarrow}^{LS}$ represent the mean escape times of carriers from a LSs into the FM/$n$-Ge on the left/right. The $\tau_{sf}$ is an (average) spin lifetime in the Ge (both LSs and Ge bulk channel) and $\tau_n$ is the (total) mean escape time from the LSs to the FM and the Ge bulk channel after creation of spin-polarized carriers at the Ge interface. $N^{LS/ch} = N_{3D}^{LS/ch} d_{LS/ch}$ are the two-dimensional density of states integrated over the thickness of the LSs layer /Ge bulk channel.

For $r_b \ll r_{LS}$, when the decoupling between the interface and the SC bulk channel by a Schottky barrier is negligible (i.e. the Schottky barrier is thin enough to facilitate the direct tunneling from a FM to SC), Eqs. (C1), (C2), and (C3) become as follows:

Single-step tunneling ($r_b \ll r_{LS}$, $r_{ch} \ll r_{LS}$),

$$\Delta\mu_{LS} \approx 2e\gamma Jr_{ch}, \quad \Delta\mu_{ch} \approx 2e\gamma Jr_{ch},$$

$$\frac{\Delta V}{V} \approx \frac{\gamma^2}{1-\gamma^2}\left(\frac{r_{ch}}{R_b^* + r_b}\right) = \frac{\gamma^2}{1-\gamma^2}\left(\frac{\tau_{sf}^{ch}}{(N^{ch}/N^{LS})(\tau_{\leftarrow}^{LS} + \tau_{\rightarrow}^{LS})}\right),$$

$$\tau_{sf} \approx \tau_{sf}^{ch}, \quad \tau_n \approx (N^{ch}/N^{LS})(\tau_{\leftarrow}^{LS} + \tau_{\rightarrow}^{LS}), \tag{C4}$$

On the other hand, for $r_b \gg r_{LS}$, when the interface is sufficiently decoupled from the SC bulk channel by a Schottky barrier (i.e. the Schottky barrier is too thick to directly tunnel from a FM to SC), Eqs. (C1), (C2), and (C3) should be considered as follows:

Two-step tunneling ($r_b \gg r_{LS}$, $r_{ch} \ll r_{LS}$),

$$\Delta\mu_{LS} \approx 2e\gamma Jr_{LS}, \quad \Delta\mu_{ch} \approx 2e\gamma J\frac{r_{LS}r_{ch}}{r_b},$$

$$\frac{\Delta V}{V} \approx \frac{\gamma^2}{1-\gamma^2}\left(\frac{r_{LS}}{R_b^* + r_b}\right) = \frac{\gamma^2}{1-\gamma^2}\left(\frac{\tau_{sf}^{LS}}{\tau_{\leftarrow}^{LS} + \tau_{\rightarrow}^{LS}}\right),$$

$$\tau_{sf} \approx \tau_{sf}^{LS}, \quad \tau_n \approx \tau_{\leftarrow}^{LS} + \tau_{\rightarrow}^{LS}, \tag{C5}$$

**FIGURE CAPTIONS**

**Figure 1.** (a) Schematic illustration of device geometry and measurement scheme. Inset: photomicrograph of the symmetric device consisting of five tunnel contacts (*a-e*) (b) Spatial distribution of the induced spin accumulations ($\Delta\mu_\pm$) by spin injection ($V_-<0$) and extraction ($V_+>0$) without/with an applied transverse magnetic field ($B_\perp$). The arrows between ($x_1$, $y_1$) and ($x_1$, $y_2$) represent the voltage drops by the tunnel contact and, the spin accumulation and part of Ge channel, respectively. (c) High-resolution TEM image of the CoFe(5 nm)/MgO(2 nm)/*n*-Ge tunnel structure. The topmost Cr layer is a capping layer to prevent oxidation of the sample. Left: low-magnification TEM image of the structure. The zone axis is parallel to the [110] direction of Ge. Middle: *in-situ* RHEED patterns of the MgO and CoFe layer along the azimuths of Ge$_{[110]}$ and Ge$_{[100]}$, respectively. Right top: SAED covering the whole region of the contact. Right bottom: simulated diffraction pattern of CoFe(001) [100] ||MgO(001)[110]||Ge(001)[100] along the [110] direction of Ge. (d) *J-V* characteristics of CoFe(5.0 nm)/MgO($t_{MgO}$=1.5, 2.0, 2.5 nm)/*n*-Ge tunnel contacts at 300 K. (e) Associated RA products (at the reverse bias voltages of -0.05, -0.15, and -0.25 V), estimated Schottky barrier heights ($\Phi_B$) and depletion regions ($W_d$) for the tunnel contacts using the conventional *I-V-T* method, respectively.

**Figure 2.** (a) Voltage changes ($\Delta V$) versus transverse magnetic field ($B_\perp$) over the temperature range 200-300 K at the bias voltages of $\mp 0.15$ V (spin injection/extraction condition) for the CoFe/MgO(2 nm)/*n*-Ge contact. (b) Voltage changes ($\Delta V$) of CoFe/Cr($t_{cr}$=0, 1.5, 3.0 nm)/MgO/Ge contacts versus transverse magnetic field ($B_\perp$) at 300 K. (c) Electrical Hanle signals ($\Delta V$) and corresponding spin RA products ($\Delta V/J$)

across the CoFe/MgO/*n*-Ge tunnel contact as a function of a transverse magnetic field ($B_\perp$) at 300 K. Data are taken with the applied current of $-14/+179$ uA, corresponding to $V_\mp = \mp 0.15$ V at $B_\perp = 0$. The solid lines represent the Lorentzian fits with $\tau_{sf,\mp} = 120/159$ ps ($V_\mp = \mp 0.15$ V). (d) Normal ($\Delta V_{\text{normal}}$) and inverted Hanle ($\Delta V_{\text{inverted}}$) effects of the contact for perpendicular ($M \perp B$, red) and in-plane ($M // B$, blue) measurement, respectively.

**Figure 3.** (a) Electrical Hanle signal ($\Delta V$), (b) spin RA product ($\Delta V / J$), and (c) TSP$^2$ ($\gamma_d \gamma_{i/e}$) with an applied bias voltage (up to $\pm 0.3$ V) over the temperature range 200-300 K. (d) Comparison of the measured spin signals ($\Delta V/J$) with the expected ones from the single-step ($r_{s\_ss}$) and two-step ($r_{s\_ts}$) tunneling process. For this calculation, we have used the representative values of $N^{LS} \sim 5 \times 10^{13}$ eV$^{-1}$ cm$^{-2}$ for MgO/Ge contact,[22] $N^{LS} \sim 1 \times 10^{14}$ eV$^{-1}$ cm$^{-2}$ for Al$_2$O$_3$/Cs/Si contact,[28] and $N^{LS} \sim 5 \times 10^{12}$ eV$^{-1}$ cm$^{-2}$ for SiO$_2$/Si contact.[37] The red, magenta, and cyan symbols represent our data, ref. 28, and ref. 19, respectively. (The closed and open triangles represent calculated spin signals from the two-step tunneling using the measured spin lifetime and optimistic value (~1 ns), respectively.) (e) Temperature dependence of magnetoresistance ($\Delta V/V$) and applied current (*I*) at the bias voltage of -0.15 V. (f) Schematic illustration for lateral inhomogeneity of tunneling current across the tunnel contact and its localization with the temperature decrease.

**Figure 4.** Structural and electrical characterizations of CoFe/Cr/MgO/Ge tunnel contacts. (a) Evolution of *in-situ* RHEED patterns during the growth processes of the CoFe(5 nm)/Cr($t_{Cr}$=0, 1.5, 3.0 nm)/MgO(2 nm)/Ge samples. The RHEED observations were carried out along the azimuths of Ge[110]. (b) *J-V* characteristics of CoFe(5 nm)/Cr($t_{Cr}$)/MgO(2 nm)/n-Ge tunnel contacts with the different Cr thickness of 0, 1.5, 2.0 nm at 300 K. (c) Arrhenius plots [$\ln(I_R/T^2)-1/T$] of the tunnel contacts with the different Cr thicknesses.

**Figure 5.** (a),(b) Schematic energy band diagrams for the CoFe/MgO/*n*-Ge tunnel contact incorporating the variation of depletion region under different bias regimes. Parabolic dispersion *E(k)* representing majority (red)/minority (blue) spin bands of ferromagnet is displaced in the energy band diagram. (c),(d) Associated spin accumulations near the *n*-Ge interface (localized states ($r_{LS}$), Ge bulk channel ($r_{ch}$)). (a)/(c) and (b)/(d) represent the forward ($V_+$>0, spin extraction) and reverse ($V_-$<0, spin injection) bias region, respectively.

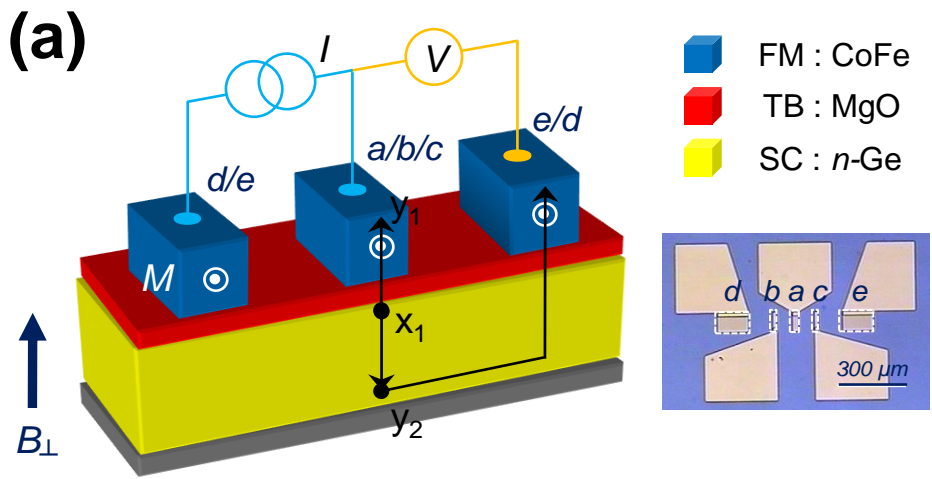
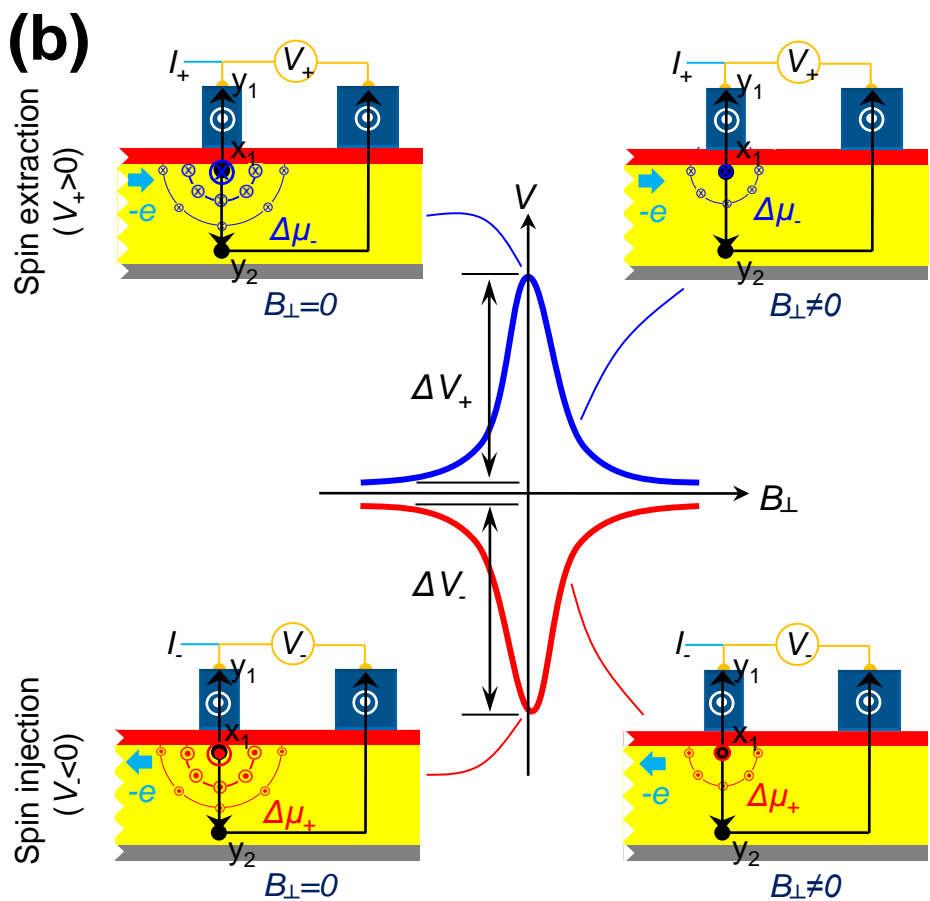
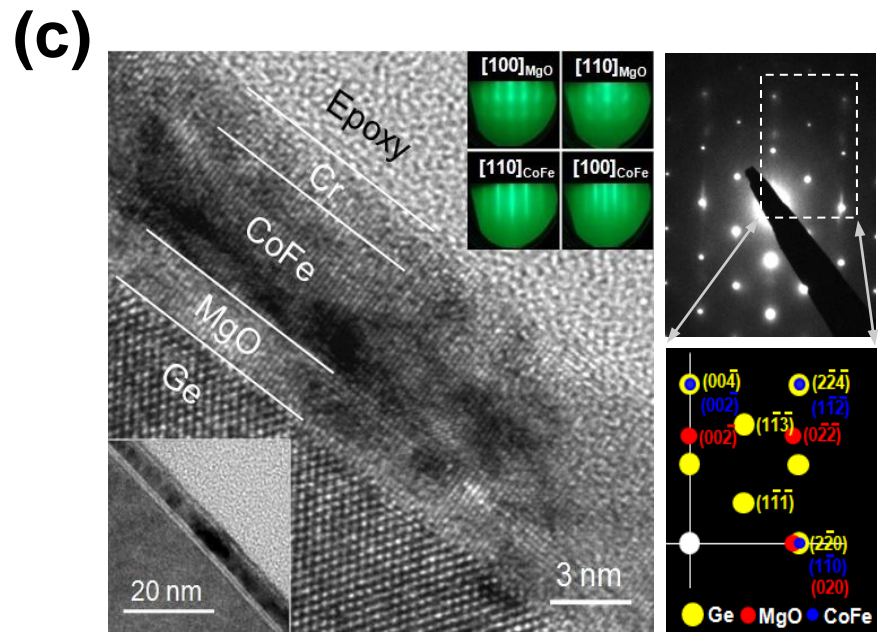
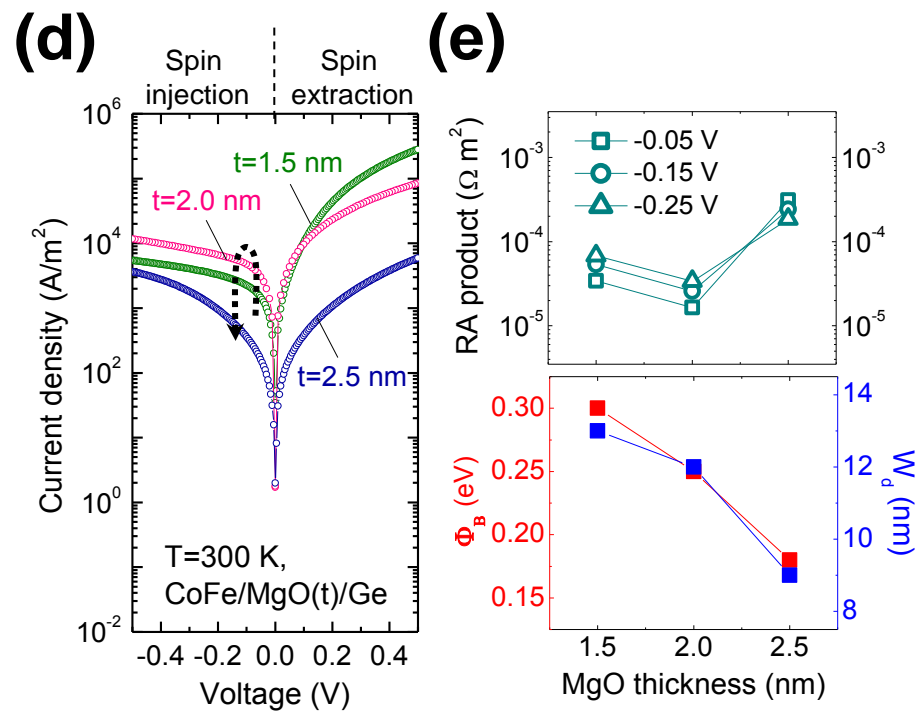

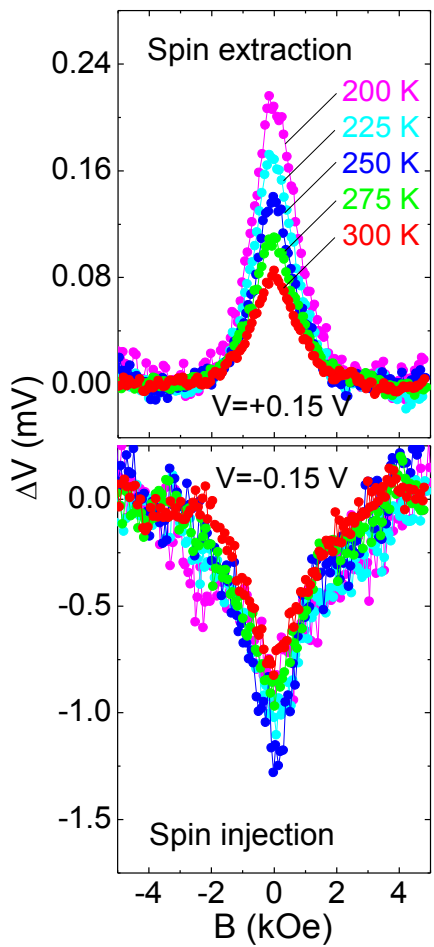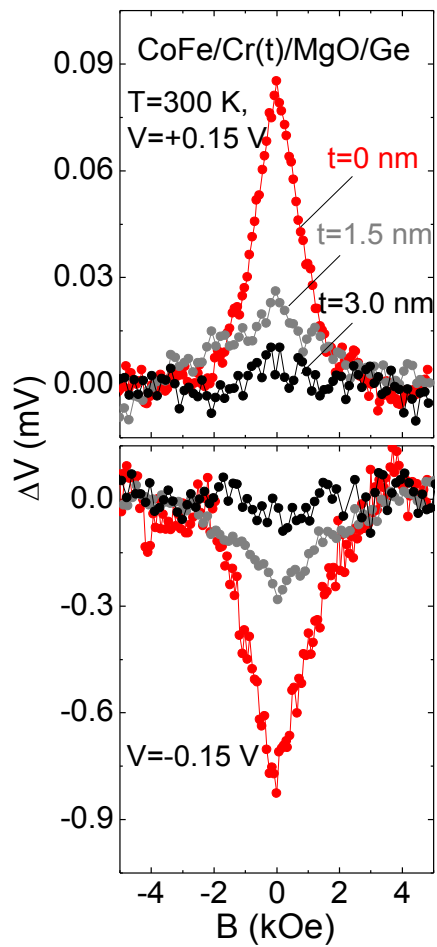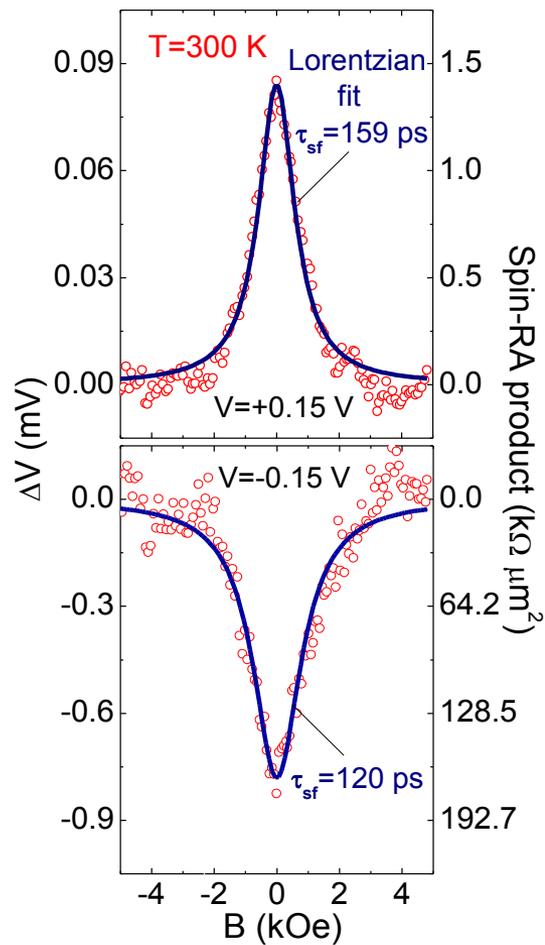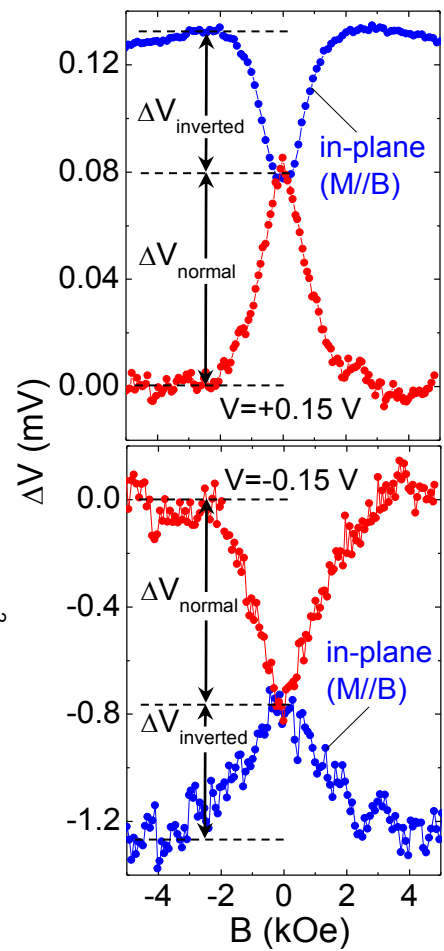

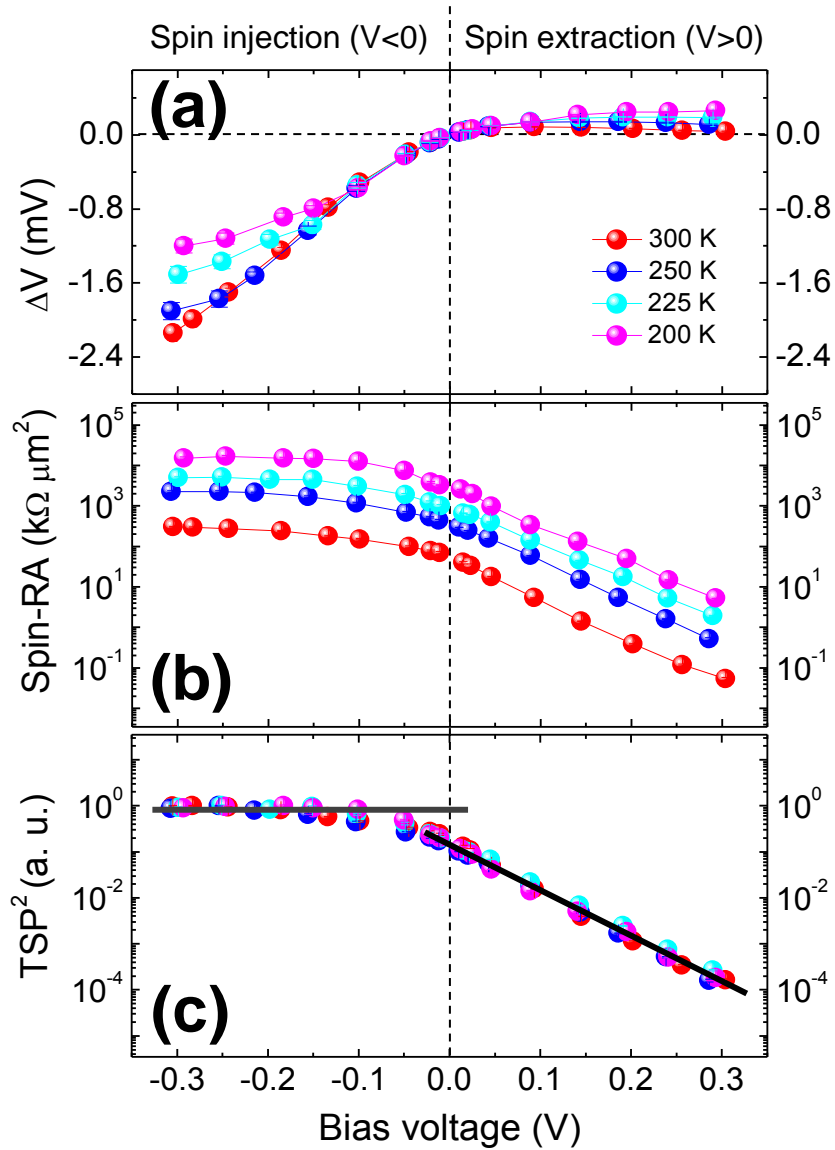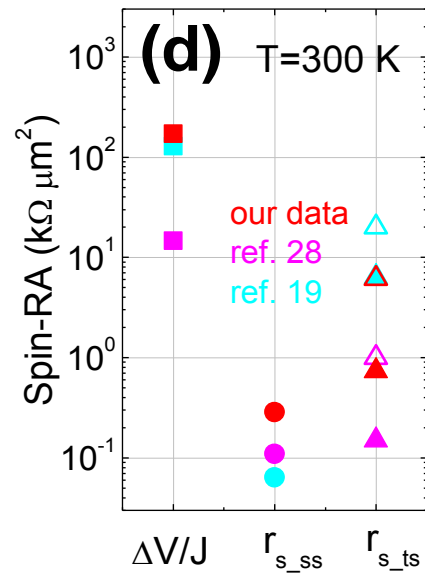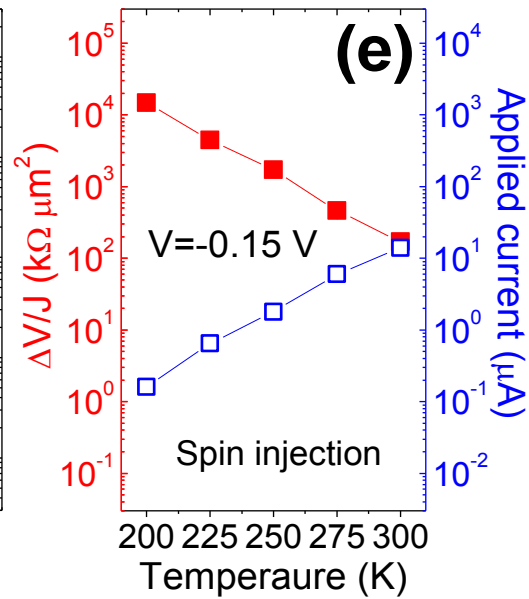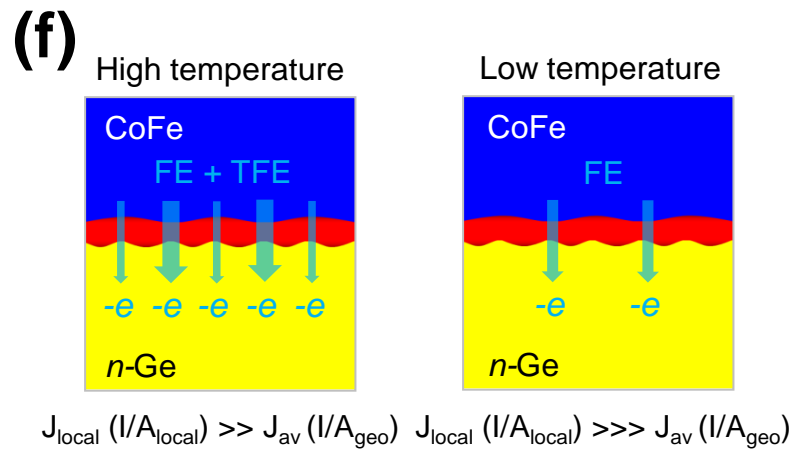

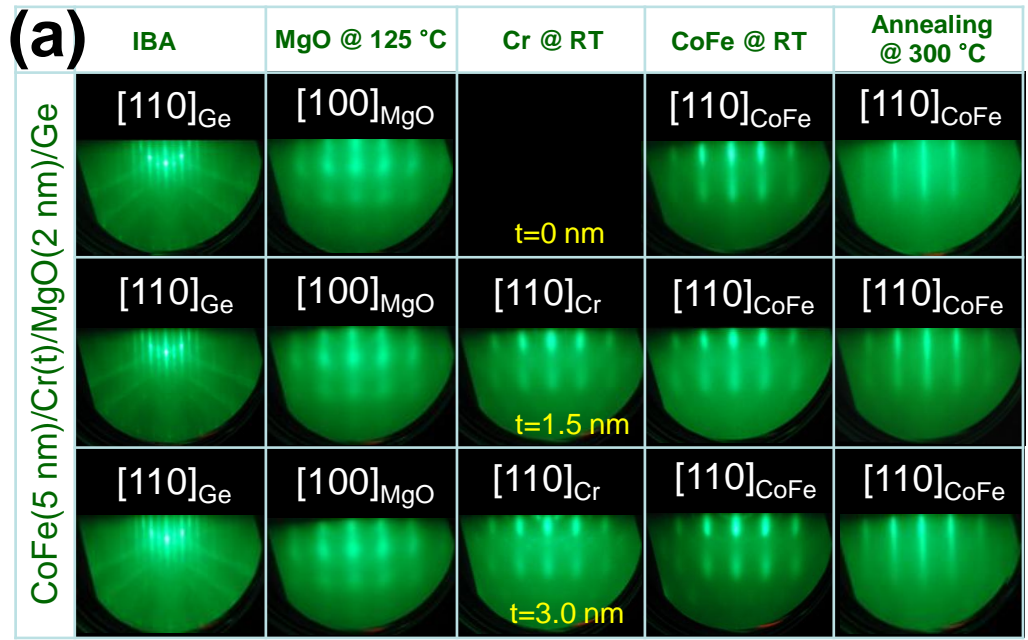
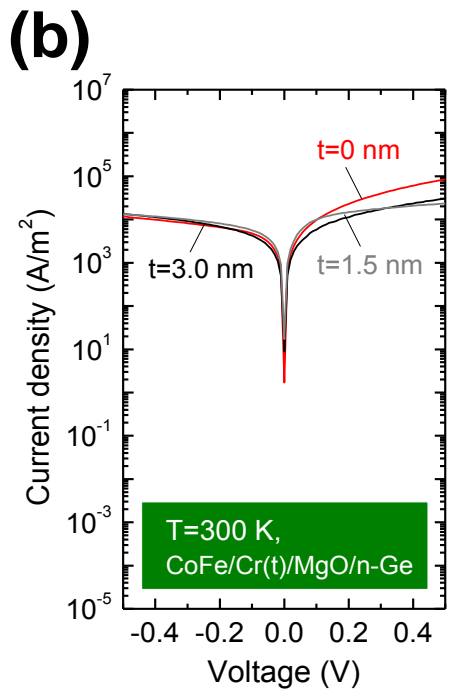
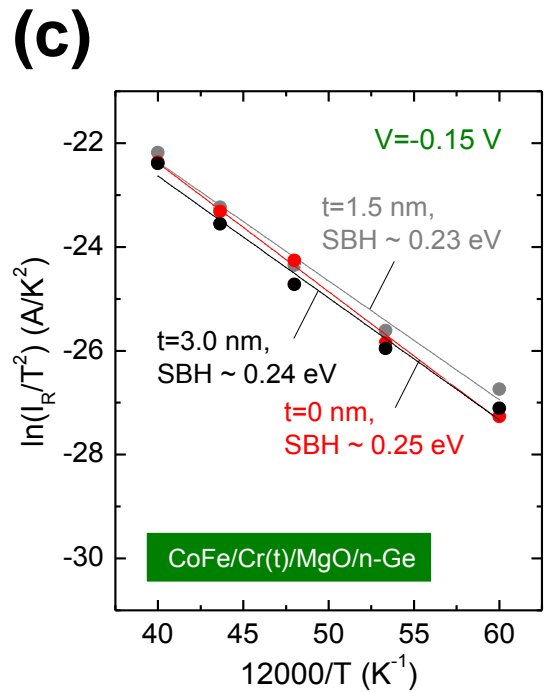

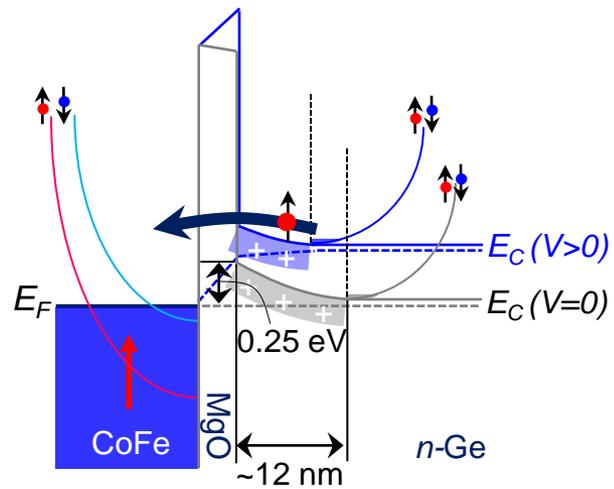
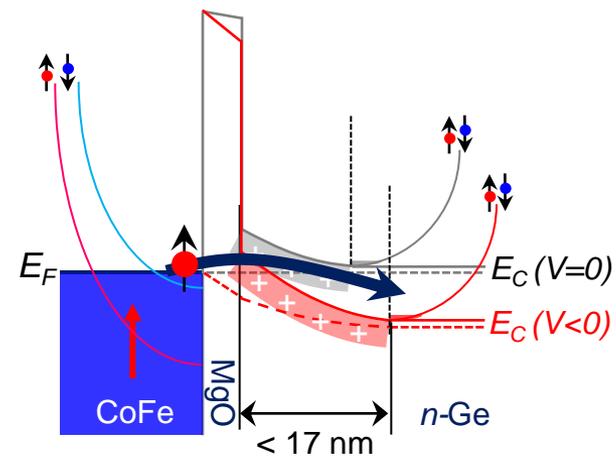
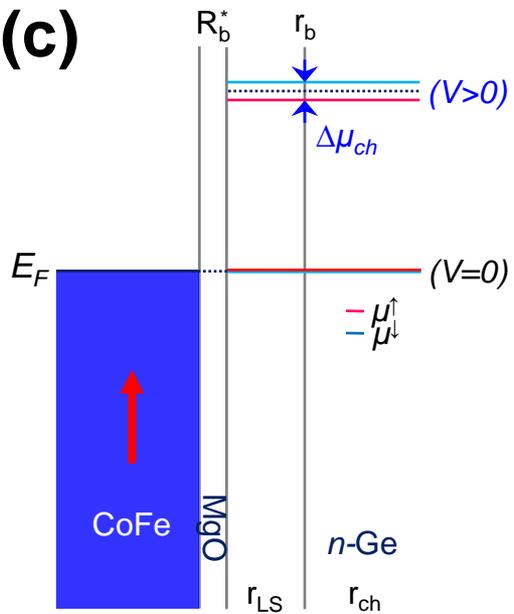
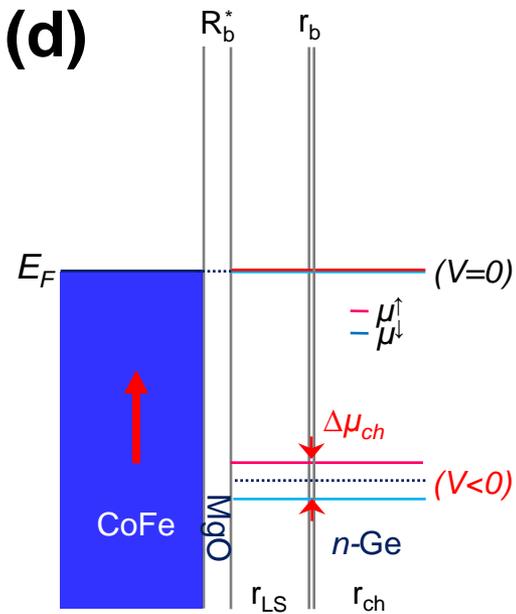